\documentclass[twocolumn,showpacs,showkeys,preprintnumbers]{revtex4-1}
\usepackage{amssymb}

\usepackage{amsfonts}

\usepackage{latexsym}

\usepackage{dcolumn}

\usepackage{amsmath}

\usepackage{graphicx}

\usepackage{psfrag,pstricks}

\begin{document}
\title{ Control and manipulation of electromagentically induced transparency in a nonlinear  optomechanical system with two movable mirrors }

\author{S.Shahidani $^{1}$ }
\email{sareh.shahidani@gmail.com}

\author{M. H. Naderi$^{1,2}$}
\author{ M. Soltanolkotabi$^{1,2}$}
\affiliation{$^{1}$ Department of Physics, Faculty of Science, University of Isfahan, Hezar Jerib, 81746-73441, Isfahan, Iran\\
$^{2}$ Quantum Optics Group, Department of Physics, Faculty of Science, University of Isfahan, Hezar Jerib, 81746-73441, Isfahan, Iran
}
\date{\today}

\begin{abstract}
We consider an optomechanical  cavity made by two moving mirrors which contains a Kerr-down conversion nonlinear crystal. We show that   the coherent oscillations of the two mechanical oscillators can  lead to splitting in   the  electromagnetically  induced transparency (EIT) resonance, and   appearance of an absorption peak within  the transparency window. In this configuration the coherent induced  splitting of  EIT  is similar to  driving a   hyperfine transition  in an  atomic Lambda-type three-level  system by a radio-frequency or microwave field. Also, we show that the presence of nonlinearity  provides an additional flexibility  for adjusting the width of the transparency windows. The combination of an additional mechanical mode and the nonlinear crystal suggests new possibilities for   adjusting the resonance frequency, the width and the spectral positions of the  EIT windows  as well as the enhancement of the absorption peak within the transparency window. 
\end{abstract} 

\maketitle

\section{Introduction}
%
%

 The coherent interaction of  laser radiation  with multi-level atoms   can induce interesting  phenomena such as electromagnetically induced transparency (EIT) and electromagnetically induced absorption (EIA).
EIT is a technique for turning an opaque medium into  a transparent one and EIA is a technique for enhancement  of absorption of light around resonance. These techniques have been used widely to manipulate the group velocity of light\cite{group1, group2, group3},  for  storage of quantum information \cite{storage1,storage2, storage3},  and for enhancement  of  nonlinear processes\cite{nonlinear1,nonlinear2,nonlinear3,nonlinear4}. 

Theoretical studies and technological advances  in nanofabrication, laser cooling and trapping \cite{kipp,favero,cooling1,rae,cooling2}  have made it possible to reach a considerable control over the light-matter interaction in an  optomechanical system. Optomechanically induced transparency (OMIT) and absorption (OMIA) are notable examples of light beam control in optomechanical systems. In OMIT which has been predicted theoretically\cite{agarwal,huang} and demonstrated experimentally\cite{weis, safavi, Karuza},   the anti-Stokes scattering of an intense red-detuned optical "control" field  brings about a  modification in   the optical response of the  optomechanical  cavity  making it transparent in a narrow bandwidth around the cavity resonance for a probe beam. 
In analogy to  the atomic EIT, the happening of OMIT  is accompanied by   a sharp negative derivative of the  dispersion profile of the cavity   near the  resonance and subluminal group velocity for the probe field \cite{tarhan, zhan}. 
In the atomic EIT the possibility of modification of the probe laser absorption, splitting and reshaping of the EIT peak and  reduction of the EIT linewidth have been studied widely\cite{modify1,modify2,modify3,modify4,modify5,modify6,modify7}.

In this work we are interested in the engineering and control of   the probe response, specially OMIT resonance, in the presence of an additional mechanical mode and the Kerr-down conversion nonlinearity. To this end, we consider a cavity with two moving mirrors,  driven by a strong coupling and a weak probe  field  which contains  a nonlinear crystal consisting of a Kerr medium and a degenerate optical parametric amplifier (OPA). This   exploration  is motivated by the following reasons.
 First, in  a recent theoretical work \cite{agarwal-EIA}  it has been shown that the coherent coupling between the two cavity modes and the mechanical mode of a moving mirror in a double cavity configuration of optomechanical system leads to the appearance of an absorption peak within the transparency window. In this  configuration    by changing the power of the electromagnetic field   the  switching between  EIT and  EIA is possible. This model is quite general and  a variety of systems, which can  be modeled by three  coupled oscillators,  can make the same  response.  A driven Fabry-Perot cavity with two vibrating mirrors  can   be effectively described by three  coupled oscillators whenever a slightly difference in the mechanical frequencies leads to   the center-of-mass-relative-motion coupling \cite{vitali1}. In this configuration the  two mechanical oscillators are coupled to a single cavity mode.
 
Second, an OPA inside a cavity can considerably improve the optomechanical coupling, the normal mode splitting (NMS), and  the cooling of the mechanical mirror\cite{agarwal-gain}.  This kind of  cooling process which is accompanied  by  the enhancement of the effective damping rate of the mirror  can be used  to increase  the width of the transparency window and reduce the  group velocity of a propagating probe pulse. 

Third,  it has been predicted\cite{kumar} that by tuning the Kerr nonlinearity in an optomechanical cavity  one can use the cavity energy shift to  reduce  the photon number fluctuation  and   provide a coherently-controlled dynamics for the mirror.

Based on these reasons,  we  first investigate the effect of  the additional mechanical oscillator on the OMIT resonance. We will show that if the coupling field  oscillates close to the mechanical resonance frequencies  there are two occasions of two-photon resonance for the  probe and coupling lasers. Consequently, the coherent oscillations of the two mechanical oscillators  give rise to splitting  of the OMIT resonance, and   appearance of an absorption peak within  the transparency window. The coherent induced  splitting of  OMIT resonance  in this configuration  is similar to  driving a   hyperfine transition  in an  atomic $\Lambda$-type three-level  system by a radio-frequency or microwave field.
 
Then we explore   how  EIT and EIA resonances  respond to the presence of a Kerr-down conversion nonlinearity in the cavity. We will show  that in the presence of Kerr-down conversion nonlinearity one can effectively control the width of the transparency window. Also, we demonstrate  that to achieve a desirable control over the  OMIT resonance the presence of   both nonlinearities is needed. 

In addition,  for the three-mode nonlinear optomechanical system we show that the coherent oscillation of the center-of-mass mode which is  responsible for the absorption peak and splitting in the transparency window, increases. This results  in the increment of the central peak absorption. 

Briefly, the combination of an additional mechanical mode and the nonlinear crystal suggests new possibilities for  "engineering"  the OMIT resonance.

\section{The Physical Model}\label{sec1}
\begin{figure}[ht]
\centering
\includegraphics[width=3.5 in]{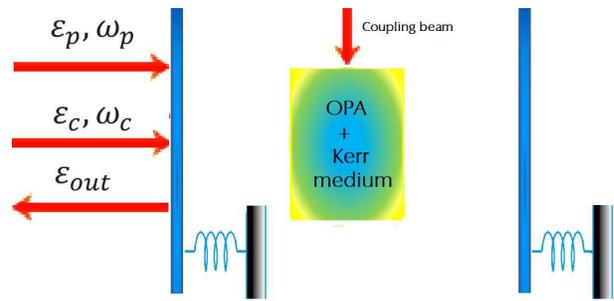}
\caption{
 (Color online) Schematic  of the setup studied in the text. The cavity that consists of two movable mirrors contains a Kerr-down conversion system which is pumped by a coupling beam to produce parametric oscillation and induce Kerr nonlinearity in the cavity. The cavity mode is coherently driven by a strong input coupling laser field and a weak probe field through the left mirror.}
\label{fig:fig1}
\end{figure}

The model we consider is an optomechanical cavity with two vibrating mirrors which contains a Kerr-down conversion nonlinear crystal (Fig.\ref{fig:fig1}). The  vibrating mirrors are  treated as two independent quantum mechanical harmonic oscillator with resonance frequency $\Omega_{k}$, effective mass $m_{k}$, and energy decay rate $\gamma_{k}$ $(k=1,2)$,  coupled to a common cavity mode having the resonance frequency $\omega_{0}$. The nonlinear crystal is composed of a degenerate  OPA  and a nonlinear Kerr medium. The cavity mode is coherently driven by a strong  input coupling laser field with frequency $\omega_{c}$ and amplitude $\varepsilon_{c}$ as well as a weak probe field with  frequency $\omega_{p}$ and amplitude $\varepsilon_{p}$   through the left mirror.  Furthermore, the system is pumped by a coupling beam to produce parametric oscillation and induce the Kerr nonlinearity in the cavity. When the detection bandwidth is chosen such that it includes only a single, isolated, mechanical resonance and mode-mode coupling is negligible we  can restrict to a single mechanical mode for each mirror  so that  the mechanical Hamiltonian of the mirrors is given by 
\begin{equation} \label{Hm}
H_{m}= \sum_{k=1}^{2}( \frac{ p_{k}^2} {2m_{k}} +\frac{1} {2} m_{k} \Omega_{k}^2 q_{k}^2).
 \end{equation}
Furthermore, in the adiabatic limit,   in which the mirror frequencies are much smaller than the cavity free spectral range $c/2L$ ($c$ is the speed of light in vacuum and $L$  is the cavity   length in the absence of the intracavity field) the photon scattering into the other modes can be neglected and we can  restrict the model to the case of single-cavity mode\cite{law, genes}. We also assume that the induced resonance frequency shift of the cavity and the nonlinear parameter of the Kerr medium are much smaller than the longitudinal-mode spacing in the cavity.
It should be noted that in the adiabatic limit, the number of photons  generated by the Casimir, retardation, and Doppler effects is negligible \cite{mancini, man-tomb,giov}.
Under this condition, the total  Hamiltonian of the system  can be written as
\begin{equation} \label{H}
H=H_{0}+H_{1},
 \end{equation}
where
\begin{subequations}\label{hamil}
\begin{eqnarray}
H_{0}&=&\hbar\omega_{0}  a^{\dagger} a+H_{m}
+\hbar g_{m} a^{\dagger} a (q_{1}-q_{2})\nonumber\\
&&+i\hbar (s_{in}(t) a^{\dagger}- s_{in}^{*}(t)a),\\  
H_{1}&=&i \hbar G( e^ { i\theta} a^{\dagger 2}  -e^ {-i\theta} a^2 )+\hbar\eta a^{\dagger 2} a^{2} .
\end{eqnarray}
\end{subequations} 
The first  term in $H_{0} $ is  the free Hamiltonian of the cavity field with the annihilation (creation) operator   $a(a^{\dagger})$, frequency $\omega_0$ and decay rate $\kappa$,  $H_m$ is the free Hamiltonian of the mirrors given by Eq.(\ref{Hm}),  the third term describes the  optomechanical coupling between the cavity field and the mechanical oscillators due to  the radiation pressure force,  and the last term in $H_{0} $  describes the driving of the intracavity mode with the input laser amplitude $s_{in}(t)$.  Also, the two terms in  $H_{1} $ describe, respectively, the coupling of the intracavity field with  the  OPA and the Kerr medium; $G$ is the nonlinear gain of the OPA which is proportional to the pump power driving amplitude, $\theta$ is the phase of the field driving the OPA, and $\eta$ is the anharmonicity parameter proportional to the  third order nonlinear susceptibility $\chi^{(3)}$  of the Kerr medium. We will solve this problem for the total driving field $s_{in}(t)=(\varepsilon_{c}+\varepsilon_{p}e^{-i(\omega_{p}-\omega_{c})t})e^{-i\omega_{c}t}$, where  $\varepsilon_{c} =\sqrt{2\kappa P_{c}/\hbar \omega_{c}}$ $(\varepsilon_{p}=\sqrt{2\kappa P_{p}/\hbar \omega_{p}})$ and $P_{c}$ ($P_{p}$) are, respectively, the amplitude and power of the input coupling (probe) field.
The dynamics of the system is described by a set of nonlinear Langevin equations. Since we are interested in the mean response of the system to the probe field we write the Langevin equations for the mean values. In a frame rotating at the coupling laser frequency $\omega_{c}$, neglecting quantum and thermal noises we obtain \begin{subequations}\label{langevin}
\begin{eqnarray}
\langle\dot{a}\rangle\; &=&-[ i (\omega_{0}-\omega_{c})+\kappa]  \langle a\rangle- i g_{m}\langle a\rangle ( \langle q_{1}\rangle- \langle q_{2}\rangle )  \nonumber\\ 
 &&-
2 i \eta \langle a^{\dagger}\rangle\langle a\rangle ^{2}+2G e^{i\theta}\langle a^{\dagger}\rangle +s_{in}(t),\\
\langle\dot{ q_{k}}\rangle &=& \langle p_{k}\rangle}/{m_{k}, \;(k=1,2),\\
\dot {\langle p_{k}\rangle}&=&-m_{k} \Omega_{k}^{2} \langle q_{k}\rangle +(-1)^{k}\hbar g_{m}\langle a^{\dagger}\rangle\langle a\rangle\nonumber\\
&&-
\gamma_{k}\langle p_{k}\rangle, \;(k=1,2).
\end{eqnarray}
\end{subequations} 
Under the assumption that the input coupling laser field is much stronger than the probe field($\varepsilon_c\gg \varepsilon_p$), we obtain the steady-state mean values of $p$, $q$ and $a$ as
\begin{subequations}\label{qsas}
\begin{eqnarray}
 p_{k}^{s}&=&0, \;(k=1,2),\\ q_{k}^{s}&=&(-1)^k\frac{\hbar g_{m}}{m_{k} \Omega_{k}^{2}}|a_{s}|^{2}\;(k=1,2), \\
a_{s}&=&\frac{\varepsilon_{c}}{\sqrt{(\Delta - 2G \sin (\theta) )^{2}+(\kappa -2G \cos (\theta)) ^{2}}},
\end{eqnarray}
\end{subequations}
where  $q_{k}^{s}$ denotes  the new equilibrium position of the movable mirrors and $\Delta=\omega_{0}-\omega_{c}+g_{m}( q_{1}^{s}-q_{2}^{s}) +2\eta |a_{s}|^{2}=\Delta_0+2\eta |a_{s}|^{2} $ is the effective detuning of the cavity which includes both  the radiation pressure  and the  Kerr medium effects. It is obvious that the optical path and hence the cavity detuning are modified in an intensity-dependent way. Since the effective detuning $\Delta $ satisfies a fifth-order equation, it  can have five real solutions and hence the system may exhibit multistability for a certain range of parameters. In our work we choose the parameters such that only one solution exists and the system has no bistability.
Now we consider the perturbation made by the probe field. The quantum Langevin equations for the fluctuations are given by
\begin{subequations}\label{langevin}
\begin{eqnarray}
\delta \dot a &=&-( i \Delta_{1}+\kappa) \delta a- i g_{m}a_{s} ( \delta q_{1}- \delta q_{2} )  \nonumber\\ 
 &&+
(2G e^{i\theta}-
2 i \eta a_{s}^{2}) \delta a^{\dagger} +s_{in}(t),\\
\delta\dot{ q_{k}} &=& \delta p_{k}}/{m_{k}, \;(k=1,2),\\
\dot {\delta p_{k}}&=&-m_{k} \Omega_{k}^{2} \delta q_{k}+
(-1)^k\hbar g_{m} a_{s}( \delta a^{\dagger}+\delta a) \nonumber\\
&&-\gamma_{k}\delta p_{k},\,(k=1,2),
\end{eqnarray}
\end{subequations} 
where $\Delta_{1}=\Delta_0+4\eta a_{s}^2$.
It is evident that the cavity mode is coupled only to the relative motion of the two mirrors, and it is therefore convenient to rewrite the above equations in terms of the fluctuations of the relative and center-of-mass coordinates:
\begin{eqnarray}
\delta Q&=&\frac{m_{1}}{M}\delta q_{1}+\frac{m_{2}}{M}\delta q_{2},\;\delta P=\delta p_{1}+\delta p_{2},\\
\delta q&=&\delta q_{2}-\delta q_{1},\; \frac{\delta p}{\mu}=\frac{\delta p_{2}}{m_{2}}-\frac{\delta p_{1}}{m_{1}},
\end{eqnarray}
where $M=m_{1}+m_{2}$ and $\mu =m_{1}m_{2}/M$ are the effective masses of the relative and center-of-mass  modes, respectively. The linearized quantum Langevin equations  for the  fluctuation operators of these coordinates  take the forms 
\begin{subequations}\label{langevin2}
\begin{eqnarray}
\langle\delta\dot{ a}\rangle &=&-( i \Delta_{1}+\kappa) \langle\delta a\rangle+ i g_{m}a_{s} \langle\delta q\rangle  \nonumber\\ 
 &&+
(2G e^{i\theta}-2 i \eta a_{s}^{2})\langle \delta a^{\dagger}\rangle +\varepsilon_{p} e^{-i(\omega_{p}-\omega_{c})t},\\
\langle\delta \dot{q}\rangle &=& \langle{\delta p}\rangle/{\mu},\\
\langle\delta\dot{ p}\rangle&=&-\mu \Omega_{r}^{2} \langle\delta q\rangle-\gamma_{r}\langle\delta p\rangle-\mu (\Omega_{2}^{2}-\Omega_{1}^{2})\langle\delta Q\rangle\nonumber\\
&&-
\frac{\mu}{M}(\gamma_{2}-\gamma_{1})\langle\delta P\rangle+
\hbar g_{m} a_{s}(\langle \delta a^{\dagger}\rangle+\langle\delta a\rangle),\\
 \langle\delta \dot{Q}\rangle &=& \langle{\delta P}\rangle/{M},\\
\langle\delta\dot{P}\rangle&=&-M\Omega_{cm}^{2}\langle \delta Q\rangle-\gamma_{cm}\langle\delta P\rangle-\mu (\Omega_{2}^{2}-\Omega_{1}^{2})\langle\delta q\rangle\nonumber\\
&&-
(\gamma_{2}-\gamma_{1})\langle\delta p\rangle,
\end{eqnarray}
\end{subequations}
where we have defined the relative motion frequency $\Omega_{r}^2=(m_{2}\Omega_{1}^2+m_{1}\Omega_{2}^2)/M$,  damping rate $\gamma_{r}=(m_{2}\gamma_{1}+m_{1}\gamma_{2})/M$ and also the  center-of-mass  frequency $\Omega_{cm}^2=(m_{1}\Omega_{1}^2+m_{2}\Omega_{2}^2)/M$ and  damping rate $\gamma_{cm}=(m_{1}\gamma_{1}+m_{2}\gamma_{2})/M$. The above equations show that even though the cavity mode interacts only with the  relative motion mode,   there is a coupling between the center-of-mass and relative motion modes when $\Omega_{1}\neq\Omega_{2}$ or  $\gamma_{1}\neq\gamma_{2}$. We will show that  the presence of this  coupling makes the  switching from EIT to EIA  possible. Now we use a  fairly standard procedure for the investigation of the probe response. Defining $\delta=\omega_{p}-\omega_{c}$, we use the following ansatz
\begin{subequations}\label{ansatz}
\begin{eqnarray}
\langle\delta a\rangle &=&A_{-}e^{-i\delta t}+A_{+}e^{i\delta t},\\
\langle\delta a^{\dagger}\rangle &=&A_{-}^{*}e^{-i\delta t}+A_{+}^{*}e^{i\delta t},\\
\langle\delta q\rangle &=&qe^{-i\delta t}+q^{*}e^{i\delta t},\\
\langle\delta Q\rangle  &=& Q e^{-i\delta t}+Q^{*}e^{i\delta t}.
\end{eqnarray}
\end{subequations}
In the original frame $A_{-}$ and $A_{+}$ oscillate  at $\omega_p$ and $2\omega_c-\omega_p$, respectively. 
Using the input-output relation\cite{in-out}, we obtain
\begin{equation}
\varepsilon_{out}+\varepsilon_{c}e^{-i\omega_c t}+\varepsilon_p e^{-i\omega_p t}=2\kappa (  a_s+\delta a)e^{-i\omega_c t}.
\end{equation}
Substituting Eq.(\ref{ansatz}) into Eq.(\ref{langevin2}) we obtain the following equations
\begin{subequations}\label{eqs}
\begin{eqnarray}
(\Theta+i\delta)A_{-}+\Gamma( A_{+})^* +i g_m a_s q+\varepsilon_p&=&0,\\
\Gamma^{*}A_{-}+(\Theta^{*}+i\delta)( A_{+})^*-i g_m a_s q&=&0,\\
\hbar g a_s(A_{-}+(A_{+})^*)+\mu\chi_{r}(\delta)q+\Lambda Q&=&0,\\
M \chi_{cm}(\delta) Q+\Lambda q&=&0,
\end{eqnarray}
\end{subequations}
where we have defined
\begin{subequations}\label{lambda-define}
\begin{eqnarray}
\Theta &=&-(\kappa+i\Delta_1),\\
\Gamma &=&2G e^{i\theta}-2i \eta a_s^2,\\
\Lambda &=&\mu(\Omega_{1}^2-\Omega_{2}^2 +i\delta (\gamma_2-\gamma_{1})),\\
\chi_{r}(\delta)\;&=&\delta^2-\Omega_{r}^2+i\delta \gamma_{r},\\
\chi_{cm}(\delta)&=&\delta^2-\Omega_{cm}^2+i\delta \gamma_{cm}.
\end{eqnarray}
\end{subequations}
From the Eq.(\ref{lambda-define}c) we find  that when $\Omega_1=\Omega_2$ and $\gamma_1=\gamma_2$,  $\Lambda=0$ and thus the center-of-mass motion is  fully decoupled  from the cavity mode and the relative motion. While whenever $\Lambda\neq0$  the  three modes are all coupled.

The total output field $\varepsilon_t$, at the  probe frequency  is given by
\begin{equation}\label{quad}
\varepsilon_t=2\kappa A_-/\varepsilon_p=\dfrac{2\kappa}{d(\delta)}\lbrace\kappa-i(\Delta_1+ \delta) -i f(\delta)\rbrace,
\end{equation}
where
\begin{subequations}\label{definf}
\begin{eqnarray}
f(\delta)&=&\hbar g_m^2 a_s^2 /\chi(\delta),
\\
\chi(\delta)&=&\mu\chi_r(\delta)-\dfrac{\Lambda^2}{M \chi_{cm}(\delta)},\\
d(\delta)&=&(\kappa-i \delta)^2+\Delta_1^2-\vert\Gamma\vert^2 \nonumber\\&&+
2 (\Delta_1+Im(\Gamma)) f(\delta).
\end{eqnarray}
\end{subequations}
The real part ($\varepsilon_R$) and imaginary part ($\varepsilon_I$)   of the field amplitude $\varepsilon_t$, respectively, show the absorptive and dispersive behavior of the output field at the  probe frequency.
These quantities can be measured by homodyne technique \cite{homodyne}. 

The structure of the output field has some main characteristics, arising from the nonlinearity  of the system and the freedom in choosing equal or unequal mechanical frequencies and damping rates.
To understand the coupling-field-induced modification of the probe response and its structure we present the results and numerical calculations in the next section.
\section{RESULTS AND DISCUSSIONS} \label{sec3}
In this section,   we first consider the bare cavity optomechanical  system  and investigate  the condition in which the coherent coupling between the mechanical and optical modes leads to OMIT and OMIA. Then we examine the effects of the Kerr-down conversion nonlinearity on these phenomena. 
\subsection{Bare cavity}
To simplify  our treatment for the bare cavity we can use the  reasonable rotating wave approximation (RWA) to  neglect the far off-resonance lower sideband ($A^{+}\simeq 0$)  in the resolved sideband regime ($\kappa\ll\Omega_k,k=1,2$)\cite{rae}. In the   resolved sideband regime  the  normal mode splitting occurs \cite{NMS1,NMS2,NMS3}. In this approximation  $\varepsilon_t$  is simplified to the following form
\begin{equation}
\varepsilon_t\simeq\dfrac{2\kappa}{\kappa+i(\Delta_0-\delta)+i\dfrac{\hbar g_m^2 a_s^2}{\chi(\delta)}}.
\end{equation}  
In what follows we investigate the two cases of equal and different mechanical frequencies  separately.
\subsubsection{Equal mechanical frequencies and damping rates ($\Lambda=0$)}
 First, we consider the case in which the  frequencies and damping rates of the two mechanical oscillators are the same, i.e., $\Omega_1=\Omega_2=\omega_m$ and $\gamma_1=\gamma_2=\gamma_m$. As mentioned before, in this condition the radiation pressure is only coupled to the   relative position of the two mirrors and the  center-of-mass becomes an isolated quantum oscillator. Therefore $\chi(\delta)=\mu \chi_r(\delta)$.
When  $\omega_p$ is close to the cavity frequency  ($\omega_p\sim\omega_0$) and  the coupling field $\omega_c$  drives the cavity on its red sideband ($\Delta_0\thicksim \omega_m$) the structure  of the resonance response of  the output field $\varepsilon_t$ is simplified to  that of a cavity with one movable mirror and effective mass $2\mu$ :
\begin{equation}
\varepsilon_t\simeq\dfrac{2\kappa}{\kappa -i x+\lbrace\beta /(\gamma_m/2-i x)\rbrace},
\end{equation}
where $\beta =\hbar g_m^2 a_s^2/2\mu$ and $x=\delta-\omega_m$ is the detuning from the line center. Therefore the   denominator of the response function is quadratic in $x$.
\subsubsection{ Different mechanical frequencies and equal damping rates ($\Lambda\neq0$)}
Now we consider the case in which the  frequency  of the mechanical oscillators is different  $\Omega_1\neq\Omega_2$ but their  damping rates are equal $\gamma_1=\gamma_2=\gamma_m$. The new aspect of  this condition is the  coupling between the center-of-mass  and the relative motion modes which results in the    anomalous EIA in the optomechanical cavity.
When  $\omega_p$ is close to the cavity frequency  ($\omega_p\sim\omega_0$) and  the coupling field $\omega_c$ is red tuned  by an amount $\omega
_m=(\Omega_1+\Omega_2)/2$  the response of the system is simplified to the following form
\begin{equation}\label{Et-OMIA}
\varepsilon_t\simeq\dfrac{2\kappa}{\kappa -i x+\dfrac{2\beta}{\delta_1 x+b_1 -\dfrac{\Lambda^2/\mu M}{\delta_2 x+b_2}}},
\end{equation}
where  $\delta_1=\omega_m+\Omega_r$, $b_1=\omega_m^2-\Omega_r^2+i \omega_m\gamma_m$ and $\delta_2=\omega_m+\Omega_{cm}$, $b_2=\omega_m^2-\Omega_{cm}^2+i \omega_m\gamma_m$. Therefore the   denominator of the  response function is cubic in $x$.

\begin{figure}[ht]
\centering
\includegraphics[width=3.5in]{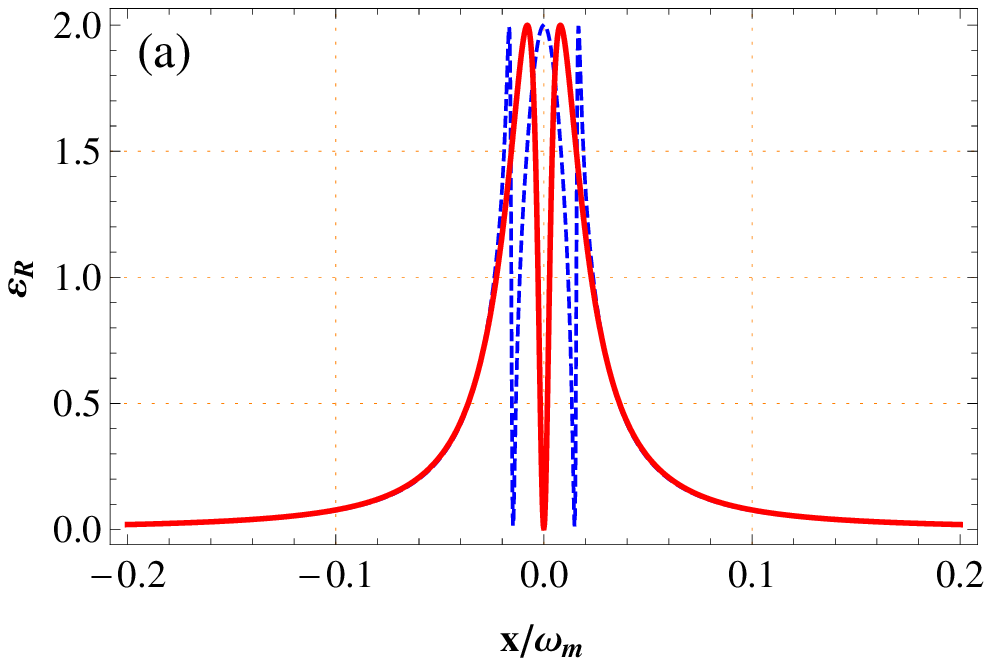}
\includegraphics[width=3.5in]{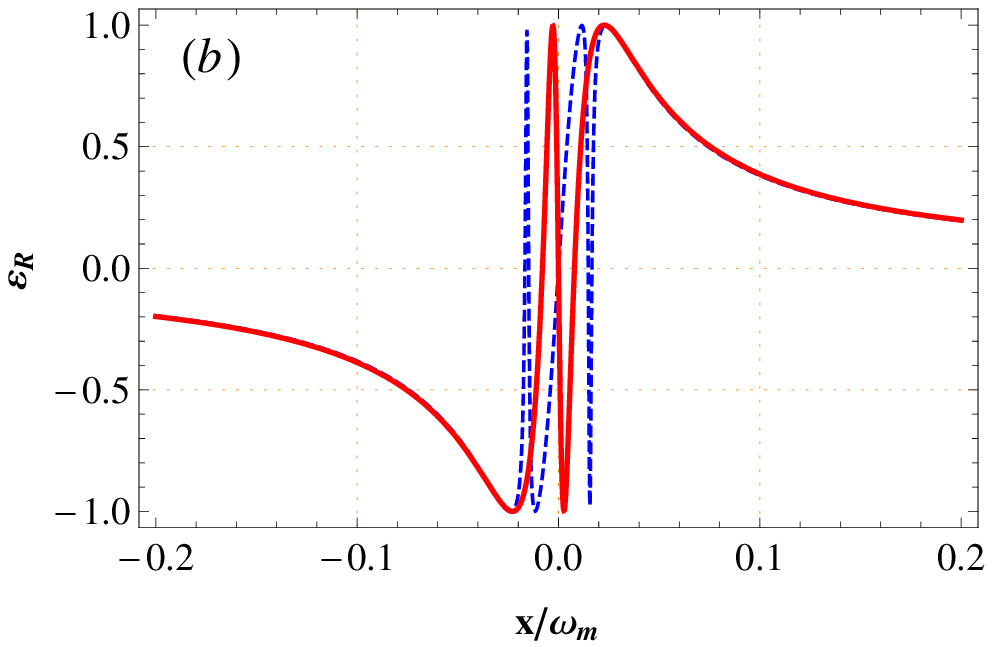}
\caption{
(Color online)(a)The real  and (b) the imaginary  parts
of the field amplitude $\varepsilon_t$ versus the normalized frequency $x/\omega_{m}$ for the  bare cavity with  equal mechanical  frequencies  $\Omega_1=\Omega_2=2\pi\times10^7$Hz(red solid line) and with different  mechanical  frequencies  $\Omega_1=2\pi\times10^7$Hz,$\Omega_2=1.03\Omega_1$ (blue dashed line).The coupling field $\omega_c$ is red detuned by an amount $\omega_m=(\Omega_1+\Omega_2)/2$ and the two mechanical damping rates $\gamma_1$ and$\gamma_2$ are equal. }
\label{fig:re&im1}
\end{figure}

To illustrate the numerical results we show the probe field absorption and dispersion profiles for the bare cavity  in Fig.\ref{fig:re&im1} for  the two cases of equal and different mechanical frequencies. We use the following set of experimentally realizable parameters \cite{teufel}: $P_c=6 $ mW,$\lambda=2\pi c/\omega_c=1064 $ nm,$\Omega_1=2\pi\times10^7$Hz, $m_1=m_2=12$ ng, $\kappa/\Omega_1=0.02 $, $\gamma_1/2\pi=\gamma_2/2\pi=200$ Hz, and  $L=6$ mm.
The figure clearly shows the splitting of the transparency window due to an additional coherency in the system.

\begin{figure}[ht]
\centering
\includegraphics[width=3in]{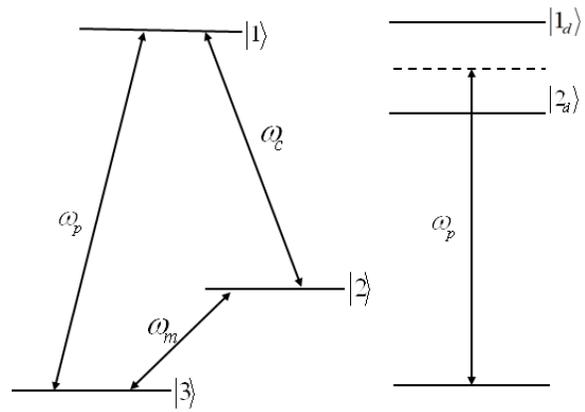}
\caption{
Level diagram structure for the OMIT. The  $\vert1\rangle\leftrightarrow \vert3\rangle$ transition is the excitation at cavity frequency and the $\vert2\rangle\leftrightarrow \vert3\rangle$ transition is the excitation of the mechanical oscillator. Coherent coupling of the mechanical and optical modes generates the destructive  interference of excitation pathways in the middle of the  doublet  of dressed states $ \vert1_d\rangle$ and $\vert2_d\rangle$ for the probe beam.}
\label{fig:level}
\end{figure}

 Physically, in the two mode optomechanical system ($\Lambda=0$) when the coupling field $\omega_c$ is red detuned by an amount $\omega_m$  ($\Delta_0\thicksim \omega_m$)   and  $\omega_p$ is close to the cavity frequency  the optomechanical system behaves like a three-level Lambda medium for the probe field as shown in Fig.(\ref{fig:level}). The intense coupling laser field   "dresses"  the mechanical mode. In this view , the OMIT can be seen as a level splitting like an Autler-Towns doublet \cite{Autler}, as shown in Fig.(\ref{fig:level}).The coherent cancellation of the two resonances in the middle of the doublet, at the two-photon resonance,  provides the system  transmittive in a narrowband around the  cavity resonance for the probe field.
\begin{figure}[ht]
\centering
\includegraphics[width=3in]{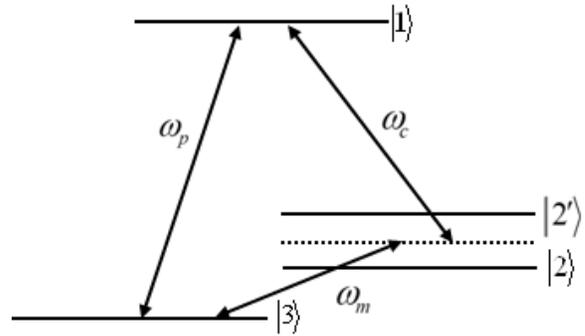}
\caption{
 Level diagram structure for the OMIA. The $\vert1\rangle\leftrightarrow \vert3\rangle$  transition is the excitation at cavity frequency; The coupling laser is red tuned by an amount  $\omega_m=(\Omega_1+\Omega_2)/2$ to induce EIT.The splitting  is due to the fact that there are two occasions of two-photon resonance for the  probe and coupling lasers.}
\label{fig:level2}
\end{figure}  

\begin{figure}[ht]
\centering
\includegraphics[width=3in]{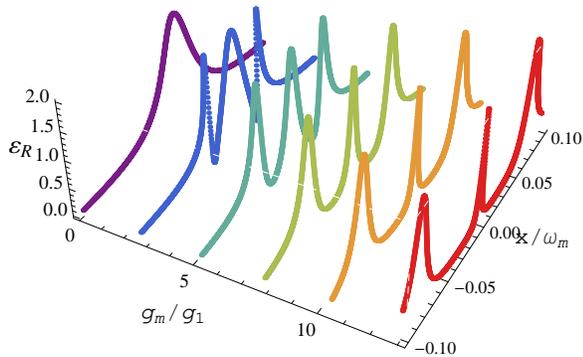}
\caption{
(Color online)The probe field absorption profile  versus the normalized frequency $x/\omega_{m}$ showing a linear OMIT splitting as a function of the  normalized radiation pressure coupling $g_m/g_1$ where $g_1=\omega_c/L$. The mechanical frequencies are $\Omega_1=2\pi\times 10^7$Hz and $\Omega_2=1.05\Omega_1$. Other parameters are the same as those in Fig.\ref{fig:re&im1}.}
\label{fig:splitting-g}
\end{figure}

Similarly, for the three-mode system ($\Lambda\neq0$) we can describe the happening of   anomalous EIA  based on a level diagram structure. In Fig.\ref{fig:level2}    the $\vert1\rangle\leftrightarrow \vert3\rangle$ transition is the excitation at cavity frequency and the coupling laser is red tuned by an amount  $\omega_m=(\Omega_1+\Omega_2)/2$ ($\Delta_0\sim\omega_m$) forming a $\Lambda$-type three-level system  producing OMIT. But the radiation pressure induces an additional coherency between the mechanical modes giving rise to OMIT splitting. The coherent induced  splitting of OMIT due to the  radiation pressure is similar to  driving a   hyperfine transition  in an  atomic $\Lambda$-type three-level  system by a radio-frequency or microwave field\cite{hyper1,hyper2,hyper3,hyper4,hyper5}. Figure \ref{fig:splitting-g} shows how the OMIT splitting varies linearly as a function of the strength of   radiation pressure coupling $g_m$.  The splitting in OMIT is due to the fact that there are two occasions of two-photon resonance for the  probe and coupling lasers at $\delta=\Omega_1$ and $\delta=\Omega_2$. 

\begin{figure}[ht]
\centering
\includegraphics[width=3in]{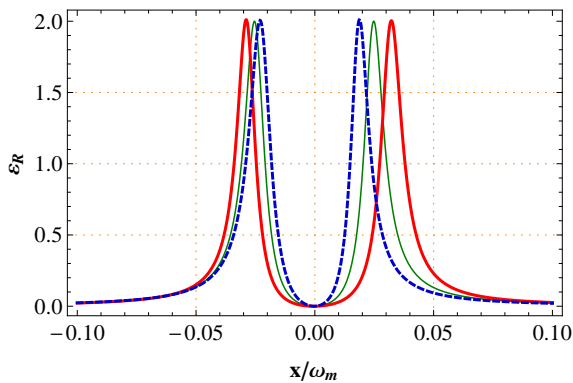}
\caption{(Color online)The absorption  profile of the  probe field versus the normalized frequency $x/\omega_{m}$ for a bare cavity ($G=\eta=0$) (green  line) and a nonlinear cavity with $G=4\times 10^6$Hz, $\eta=0.03$Hz, $\theta=3\pi/2$(blue dashed line) and with $G=4\times 10^6$Hz, $\eta=0.04$Hz, $\theta=\pi/2$(red solid line) . The parameters  are $P_c=8$mW, $m_1=m_2=15$ ng,  $\Omega_1=\Omega_2=2\pi\times 10^7$Hz, $\lambda=512$ nm, $L=2$mm, and  $\kappa=0.01\Omega_1$. Other parameters are the same as those in Fig.\ref{fig:re&im1}.
}
\label{fig:EIT-control}
\end{figure}

\begin{figure}[ht]
\centering
\includegraphics[width=3in]{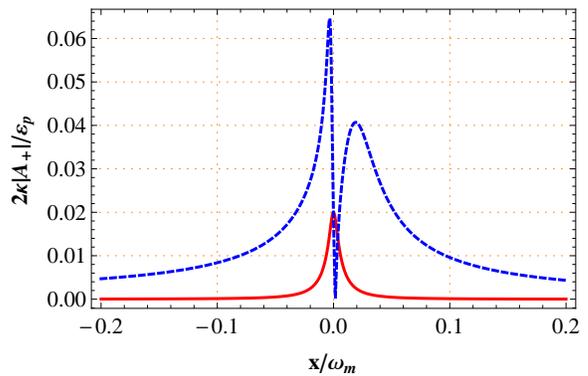}
\caption{
(Color online)The parameter  $2\kappa \vert A_+\vert/\varepsilon_p$ versus the normalized frequency $x/\omega_{m}$ for a bare cavity ($G=\eta=0$) (red solid line) and a nonlinear cavity ($G=1.5\kappa$, $\eta=0.03$Hz,$\theta=\pi/2$)(blue dashed line). The mechanical frequencies are $\Omega_1=\Omega_2=2\pi\times 10^7$Hz. Other parameters are the same as those in Fig.\ref{fig:re&im1}.}
\label{fig:sideband}
\end{figure} 
\subsection{Nonlinear cavity}
Now we investigate the effect of the  Kerr-down conversion nonlinearity on the total output field amplitude $\varepsilon_t$.  Although the nonlinearity does not alter the level diagram structure of the OMIT, it     manifests itself  in the steady-state response of the system (Eq.(\ref{qsas})), in the optomechanical coupling rate $g_m a_s$ (Eq.(\ref{langevin})), and   in the  parameter $\Gamma_+$ which is responsible for a direct coupling between $A_{-}$ and $A_+$ (Eq.(\ref{eqs}a)).

In the OMIT condition   the optomechanical coupling rate $g_m a_s$   is equivalent to the Rabi frequency   in the atomic EIT\cite{weis}. The dependence  of $ a_s$  on the nonlinearity can be used to control the width of the transparency window which is related to the effective mechanical damping  rate  $\gamma_{eff}$.
 This parameter  is approximately given by\cite{weis,safavi,agarwalrouter}
 \begin{equation}
 \gamma_{eff}=\gamma_m(1+C),
 \end{equation} 
 where $C=2 \hbar( g_m a_s)^2/m \omega_m\kappa \gamma_m$   denotes the optomechanical cooperativity of the cavity\cite{weis,rae,hill}.  In Fig. \ref{fig:EIT-control} we have plotted the absorption profile for different values of $G$, $\theta$ and $\eta$.  It shows that by controlling these parameters  the width of the  transparency window  can be increased or decreased in comparison with that of  a bare cavity. 
 It should be noted  that in the presence of only one of the two  nonlinearities we cannot control the transparency window desirably. This can be explained by the fact that according to Eq.(\ref{quad}), in the absence of optomechanical coupling ($g_m=0$) there would be an absorption peak near the modified resonance condition of the cavity $\delta=\sqrt{\Delta_1^2-\vert\Gamma\vert^2}$. Therefore the nonlinear parameters should be  choosen such that $\sqrt{\Delta_1^2-\vert\Gamma\vert^2}\simeq\Delta$, otherwise  the control and probe fields induce a radiation-pressure force oscillating at the frequency $\delta$, which is not   close enough to the resonance frequency of the moving mirrors to induce coherent oscillations in them.   This feature  leads to disappearance of OMIT in the output probe field. 
\begin{figure}[ht]
\centering
\includegraphics[width=3in]{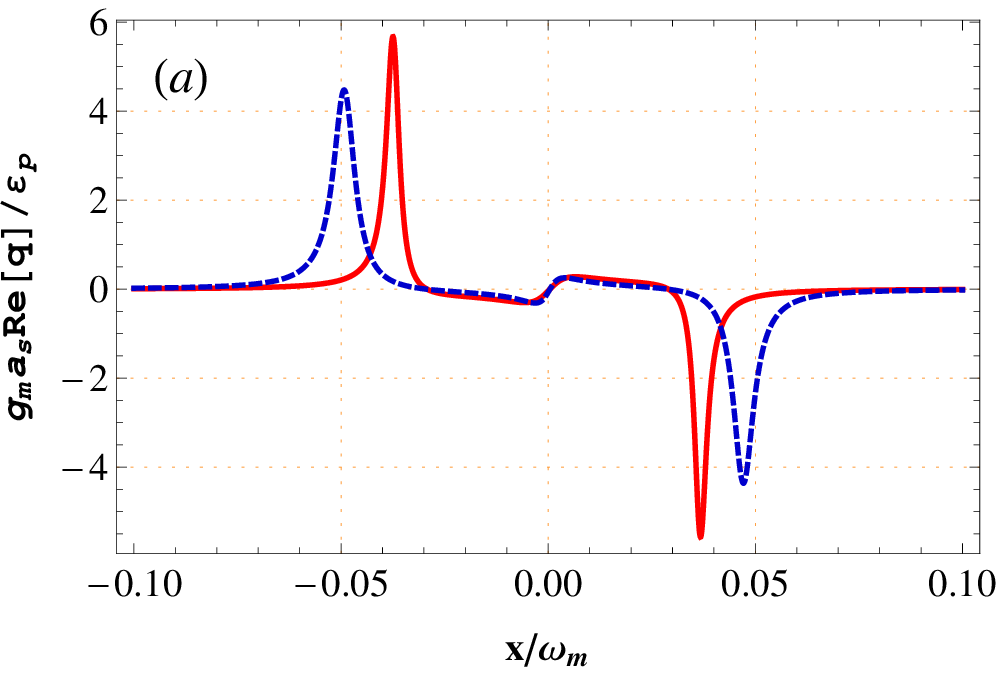}
\includegraphics[width=3in]{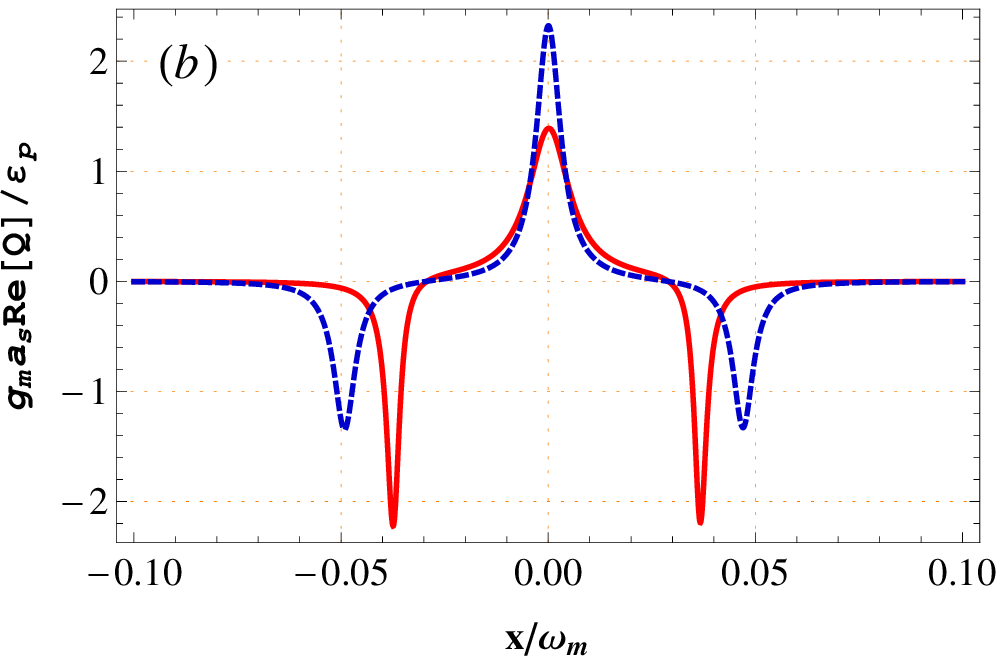}
\includegraphics[width=3in]{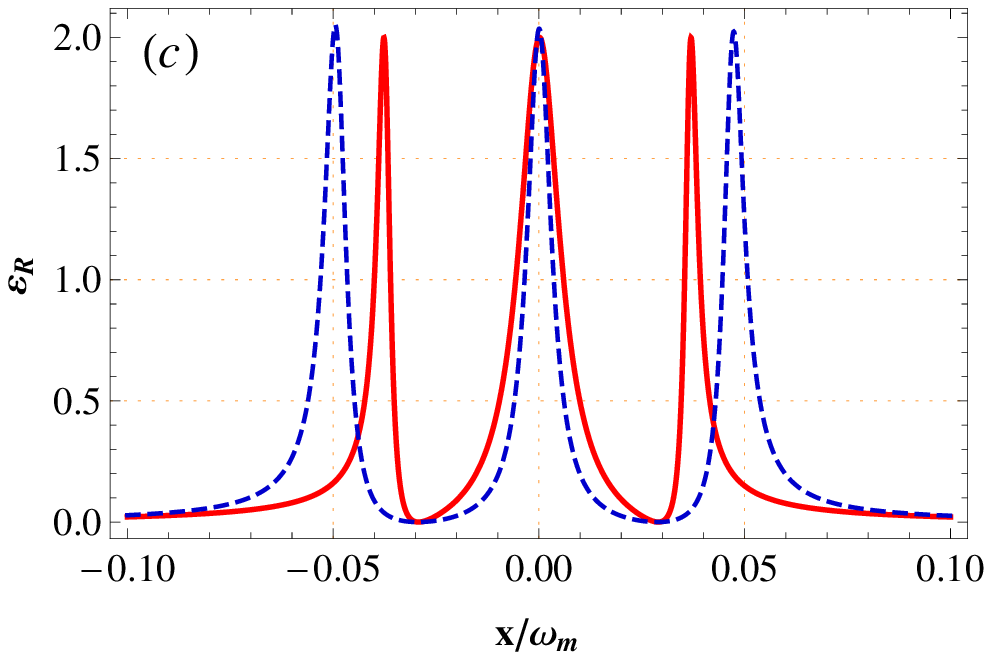}
\caption{(Color online)The real parts of (a) the normalized parameter $g_{m}a_{s} q/\varepsilon_p $, (b) the normalized parameter $g_{m}a_{s}  Q/\varepsilon_p$ and (c) the field amplitude $\varepsilon_t$ versus the normalized frequency $x/\omega_{m}$ for a bare cavity ($G=\eta=0$) (red  line) and a nonlinear cavity with $G=10^7$Hz, $\eta=0.09$Hz, $\theta=\pi/2$(blue dashed line).The mechanical frequencies are $\Omega_1=2\pi\times 10^7$Hz and $\Omega_2=1.06\Omega_1$. Other parameters are the same as those in Fig.\ref{fig:EIT-control}.
}
\label{fig:q1}
\end{figure} 

Also, according to Eq.(\ref{eqs}a), in the presence of nonlinearity there is a direct coupling between  $A_{-}$ and $A_+$ because of the factor $\Gamma$.  Therefore it seems that in contrast to the bare optomechanical cavity the Stokes scattering of the light from the strong intracavity coupling field is no longer negligible.   In Fig.\ref{fig:sideband} we have plotted the parameter $2\kappa \vert A_+\vert/\varepsilon_p$ as a function of the normalized frequency $x/\omega_{m}$ for a bare cavity and a cavity with Kerr-down conversion nonlinearity. As is seen,   in the dip of the transparency  window  $2\kappa \vert A_+\vert/\varepsilon_p$ reaches its  local minimum  for a nonlinear cavity and  its local maximum for a bare cavity. Hence even though in the presence of nonlinearity  outside the OMIT window the lower sideband    can also be tuned by the strong coupling field, but  the contribution of the Stokes scattering  around the cavity resonance  is more negligible for a nonlinear cavity.

 Now we consider the probe response in the presence of Kerr-down conversion nonlinearity for the second case ($\Omega_1\neq \Omega_2$ ). As stated before, the OMIT splitting and appearance of the central absorption peak  are   due to an additional coherent oscillation in the system which is provided by the fluctuations in the center-of-mass mode, i.e., $\langle \delta Q\rangle$.  Figures \ref{fig:q1}(a) and \ref{fig:q1}(b)  illustrate  the effect of the nonlinearity on  $q$ and $Q$, respectively. They show a shift in the coherent oscillations of $q$ which leads to the broadening of the width of the transparency windows and  an increase in   the coherent oscillations of the center-of-mass mode  $Q$  which results in the enhancement of   central peak absorption (Fig.\ref{fig:q1}(c)).

 In conclusion, we  have studied theoretically the effect of an additional mechanical mode and a Kerr-down conversion nonlinear crystal on the EIT resonance in an optomechanical system with two movable mirrors. We have  shown  that   the coherent oscillations of the two mechanical oscillators can  lead to splitting in   the  EIT resonance, and   appearance of an absorption peak within  the transparency window. This configuration  is similar to  driving a   hyperfine transition  in an  atomic $\Lambda$-type three-level  system by a radio-frequency or microwave field. Also, we have shown that in the presence of Kerr-down conversion nonlinearity    by controlling the nonlinear  parameters   $G$, $\eta$  and $\theta$ the width of transparency can be adjusted to be  greater or smaller than that of  a bare cavity. The combination of an additional mechanical mode and nonlinear crystal suggests new possibilities for manipulating and controlling the EIT resonance in the optomechanical systems.
  \section*{Acknowledgement}
The authors wish to thank The Office of Graduate Studies of The University of Isfahan for their support.
\bibliographystyle{apsrev4-1}

\end{document}